\documentclass[pdflatex,sn-mathphys-num]{sn-jnl}
\usepackage{graphicx}
\usepackage{multirow}
\usepackage{amsmath,amssymb,amsfonts}
\usepackage{amsthm}
\usepackage{mathrsfs}
\usepackage[title]{appendix}
\usepackage{xcolor}
\usepackage{textcomp}
\usepackage{manyfoot}
\usepackage{booktabs}
\usepackage{algorithm}
\usepackage{orcidlink}
\usepackage{algorithmicx}
\usepackage{algpseudocode}
\usepackage{listings}

\theoremstyle{thmstyleone}

\theoremstyle{thmstyletwo}

\theoremstyle{thmstylethree}

\graphicspath{{PLOTS/}}

\begin{document}

\title[Discovery of PISN Mass Gap from GWTC-4]{Binary Black Hole Phase Space Discovers the Signature of Pair Instability Supernovae Mass Gap}

\author*[1]{\fnm{Samsuzzaman} \sur{Afroz}\orcidlink{0009-0004-4459-2981}}\email{samsuzzaman.afroz@tifr.res.in}
\author*[1]{\fnm{Suvodip} \sur{Mukherjee}\orcidlink{0000-0002-3373-5236}}\email{suvodip@tifr.res.in}

\affil*[1]{\orgdiv{Department of Astronomy and Astrophysics}, \orgname{, Tata Institute of Fundamental Research}, \orgaddress{\street{Dr Homi Bhabha Road}, \city{Mumbai}, \postcode{400005}, \state{Maharashtra}, \country{India}}}

\abstract{The rapidly expanding catalog of gravitational-wave detections provides a powerful probe of the formation history of compact binaries across cosmic time. In this work, we extend the Binary Compact Object (BCO) phase-space framework to the full set of events in the GWTC-4 catalog to map the observed binary formation scenarios in a data-driven way. Applying this framework, we identify distinct regions of phase-space associated with different channels and discover for the first time a unique mass-cutoff scale in a data-driven way. The mapping of these on different formation channels reveals a population of first-generation (1G) black holes sharply truncated at approximately 45.5 $M_\odot$, consistent with the theoretically predicted pair-instability supernova (PISN) mass gap. These findings demonstrate the capability of the BCO phase-space to disentangle overlapping formation pathways, establish robust connections between gravitational-wave observations and binary evolution, and highlight the potential of upcoming observing runs to reveal rare populations and exotic origins.}

\maketitle

\section{Introduction}

The discovery of gravitational waves (GWs) has revolutionized our ability to investigate compact objects, providing direct access to the dynamics of strongly curved space-time and the astrophysical environments in which binary systems evolve \cite{Bailes:2021tot, Arimoto:2021cwc,Mapelli:2021taw,Iorio:2022sgz,Barrett:2017fcw,Dominik:2012kk,Bailyn:1997xt}. Since the first binary black hole (BBH) merger was reported by LIGO in 2015, subsequent observing runs by the LIGO\cite{LIGOScientific:2016dsl}-Virgo\cite{VIRGO:2014yos}-KAGRA\cite{KAGRA:2020tym} (LVK) collaboration have rapidly expanded the catalog of GW events, now including not only BBHs, but also neutron star–black hole (NSBH) binaries and binary neutron stars (BNSs). Each new detection enriches our ability to explore fundamental questions about the life cycle of massive stars, the role of stellar dynamics, and the possible contribution of non-stellar origins such as primordial black holes to the present-day compact binary population.

Traditional approaches to population inference typically rely on hierarchical Bayesian analyses, where posterior samples from individual events are combined to constrain global population parameters such as the merger-rate density, mass distribution, and spin distribution of compact binaries. These studies have provided valuable insight into the statistical properties of the detected sources and their redshift evolution \cite{LIGOScientific:2018mvr,LIGOScientific:2016aoc,KAGRA:2021duu,Bouffanais:2021wcr,Franciolini:2021tla,Cheng:2023ddt,Antonelli:2023gpu,2023ApJ...950..181W,Tiwari:2020otp,Tiwari:2021yvr, Tiwari:2025lit, Kimball:2020opk, Kimball:2020qyd}. 

However, such methods are not naturally designed to disentangle the relative contribution of multiple formation channels. The core astrophysical challenge remains: how can we robustly connect the observed GW signatures of binary compact objects (BCOs) to their underlying formation histories?

One promising route to address this question is to construct a representation that explicitly links the observable parameter space of GW detections to the evolutionary pathways predicted by different formation scenarios. Recently, we introduced the concept of the Binary Compact Object (BCO) Phase Space \cite{Afroz:2024fzp,Afroz:2025efn}, a framework that embeds GW observables such as component masses, spins, and luminosity distances into a space where trajectories corresponding to distinct formation channels can be mapped and compared. In this formulation, isolated binary evolution, dynamical assembly in dense stellar environments, mergers in active galactic nucleus (AGN) disks, and primordial black hole scenarios are each expected to trace different evolutionary tracks across the phase space \cite{Miller:2008yw,Banerjee:2010,Samsing:2014, Fragione:2023kqv,Rodriguez:2016,Zevin:2021,Bird:2016dcv,Clesse:2016vqa,Sasaki:2016jop}. 

The phase-space approach starts from the observed posteriors of individual events and embeds them directly into the multidimensional space of masses, spins, and distances. Within this space, each formation scenario is represented by a predicted “trajectory”, a physically motivated distribution that describes how binaries from that channel are expected to populate the observable parameter space as a function of redshift. The comparison between data and theory therefore becomes a matter of quantifying overlaps between observed event posteriors and these channel trajectories. This enables us to ask, for each event, which trajectories it is most consistent with, and for the population as a whole, how much of the observed phase space is covered by each channel. In this sense, the disentangling is explicit: different channels predict distinct geometries in the phase space, and the degree of overlap with observed data provides a direct measure of their relative contribution. This has two important consequences. First, it provides a transparent, event-by-event way to evaluate the consistency of events with different channels, rather than only assigning overall weights to models. Second, because the mapping is geometric, the method can naturally highlight regions of phase space where the data are not explained by existing channels, pointing toward the possibility of previously unanticipated populations. In this way, the phase-space framework complements existing population analyses by offering a different and more direct perspective on the astrophysical interpretation of GW data. Moreover, the phase-space demonstration of astrophysical population of observed compact objects can also capture any known formation channel, which can appear as a new trajectory in the phase-space volume, and non-overlapping with the theoretically known models. Some of the other well-known phase-space demonstration of the astrophysical observations are color-magnitude diagram \cite{1914PA.....22..275R}, mass and velocity dispersion diagram \cite{2002ApJ...574..740T}, the phase diagram of QCD \cite{Orsaria:2019ftf}, etc.

The initial application of this framework to the GWTC-3 \cite{KAGRA:2021duu} catalog demonstrated its ability to reveal such distinctions, identifying subsets of BBH mergers that align with trajectories expected from hierarchical mergers or AGN-assisted growth, while also suggesting possible overlap with primordial channels \cite{Afroz:2024fzp}. A subsequent extension of the method to low-mass compact objects showed that binaries straddling the neutron star-black hole mass boundary occupy particularly informative regions of phase space, with implications for understanding the putative “lower mass gap” and the possible existence of exotic populations \cite{Afroz:2025efn}. These earlier studies were necessarily limited by the number of events available, which constrained the scope of formation channels that could be considered in detail.

The landscape has changed significantly with the release of the GWTC-4 \cite{LIGOScientific:2025slb} catalog, which adds 86 confident detections to the 90 events reported in GWTC-3 \cite{KAGRA:2021duu}, bringing the total number of well-characterized events to 176 when combined with those from earlier observing runs. This expanded dataset spans a broader range of binary configurations, from nearly equal-mass to asymmetric mass ratio compact objects. The increase in both sample size and diversity makes it possible to undertake, for the first time, a comprehensive phase-space analysis of the entire LVK catalog to date, enabling a unified exploration of formation pathways across all classes of compact binaries.

In this work, we apply the BCO Phase Space framework to the complete GWTC-4 catalog, with the exception of GW170817 \cite{LIGOScientific:2017vwq}, which is a confirmed binary neutron star identified through both its mass properties and the detection of electromagnetic counterparts. Our analysis shows that the enlarged catalog occupies well-defined regions of phase space that align with the evolutionary trajectories predicted by different formation channels. 

\begin{figure}
\centering
\includegraphics[width =12cm, height=5.0cm]{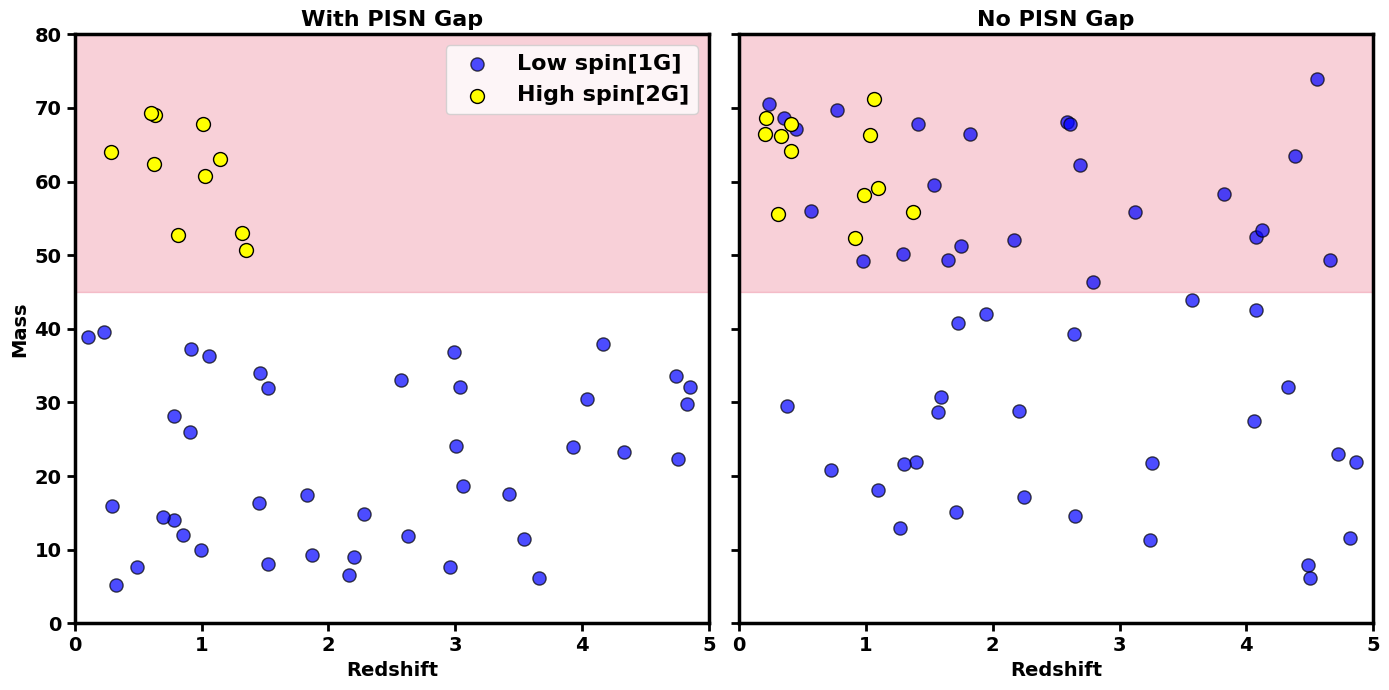}
\caption{Schematic illustration of binary black hole populations in mass-redshift space under two scenarios. The left panel shows the presence of the PISN mass gap (shaded red region), where only high-spin, high-mass second-generation (2G) black holes (yellow points) populate the mass gap at low redshift due to merger time delays, while low-spin, low-mass first-generation (1G) black holes (blue points) are distributed across all redshifts below the mass gap. The right panel shows the absence of the PISN gap, with low-spin black holes populating the full mass and redshift range, and high-spin, high-mass black holes concentrated at lower redshifts.}
\label{fig:Motivation}
\end{figure}

Crucially, we identify a distinct boundary in the black hole mass distribution that aligns with the theoretically predicted upper limit imposed by pair-instability supernova (PISN) processes. This mass scale, typically situated around 40\text{-}50 solar masses, marks the threshold beyond which first-generation (1G) black holes formed via stellar collapse are strongly suppressed due to the explosive disruption of massive stellar progenitors \cite{Farrell:2020zju,Belczynski:2019fed,Spera:2017fyx,Farmer:2020xne,Hendriks:2023yrw,Farmer:2019jed,Leung:2019fgj}. By projecting GW populations onto the phase space, this PISN mass gap emerges naturally as a defining feature, offering critical insight into the astrophysical mechanisms governing black hole formation. This enables clear differentiation between black holes formed from isolated stellar evolution and those arising from hierarchical mergers or exotic channels. Leveraging the expanded GWTC-4 catalog, this refined phase-space method provides a powerful and incisive tool to disentangle the diverse origins of black holes observed through gravitational waves, advancing our understanding of stellar evolution and compact object formation.

This physical constraint is visually illustrated in Figure~\ref{fig:Motivation}, which schematically depicts the distinct distributions of 1G and 2G black holes in mass-redshift space. The figure highlights how 1G black holes are confined below the PISN mass threshold across a broad range of redshifts, while 2G black holes populate the high-mass regime predominantly at lower redshift due to merger time delays. These evolutionary trajectories emphasize the natural emergence of the PISN gap and motivate our phase-space approach.

\section{Characterizing the Formation Pathways for Astrophysical Black Holes in Phase space} 
\label{sec:PhaseSpaceModel}

Astrophysical black holes arise from the collapse of massive stars and represent the dominant population in current GW observations. Their properties such as mass distribution and spin evolution depend sensitively on the underlying formation mechanisms, making them a valuable probe of stellar and dynamical astrophysics. Several formation channels have been proposed in the literature, including isolated binary evolution, hierarchical mergers, dynamical interactions in dense stellar clusters, and migration and capture within active galactic nucleus (AGN) disks. Each of these channels imprints distinct signatures on the resulting black hole population, influencing merger rates, spin distributions, and mass spectra.

In this work, however, we restrict our attention to two primary pathways: isolated binary evolution and hierarchical mergers. The isolated channel provides a baseline scenario where black holes originate from the evolution of massive stellar binaries through processes such as mass transfer and common-envelope evolution. In contrast, hierarchical mergers, typically occurring in dense environments, naturally lead to heavier and potentially rapidly spinning remnants through repeated mergers.

By focusing on these two channels, we capture the essential contrast between “field” binaries shaped by stellar evolution and merger-driven growth in dynamical environments, while leaving the inclusion of additional channels to future, more data-driven studies.

\subsection{Isolated Binary} 

In the isolated binary evolution scenario, black holes originate from the core collapse of massive stars within binary systems, with minimal influence from external dynamical interactions. This channel primarily forms 1G black holes, whose masses are governed by stellar evolution. A key feature is the pair-instability supernova (PISN) mechanism, which is expected to set a lower cutoff of about $40$-$50~M_\odot$ for the masses of 1G black holes \cite{Farmer:2020xne,Vink:2024dgm}. Nevertheless, this limit is subject to uncertainties linked to the treatment of fallback, angular momentum redistribution during collapse, and metallicity effects, since lower-metallicity stars lose less mass through winds and therefore can yield heavier black holes \cite{Zhang:2004kx,Mandel:2018hfr}.  

Furthermore, the binary’s evolutionary pathway, including episodes of mass transfer and possible common-envelope phases, has a strong impact on both the resulting black hole mass spectrum and their spin distributions \cite{Marchant:2021hiv, Bavera:2020uch, Mapelli:2021taw, Bavera:2020inc}.  

The cosmic merger rate of such systems is typically modeled by combining the Madau-Dickinson star formation history with an assumed delay-time distribution \cite{Madau:2014bja,Dominik:2014yma,Karathanasis:2022hrb,Mukherjee:2021rtw,Karathanasis:2022rtr}. The delay time ($t_d$) denotes the interval between the birth of stars that eventually collapse into black holes and the merger of the resulting binary black holes. Crucially, this is not a single fixed timescale but follows a distribution that captures the diversity of possible evolutionary pathways. This distribution is commonly expressed as \cite{Dominik:2014yma,Karathanasis:2022hrb,Mukherjee:2021rtw,Karathanasis:2022rtr}:

\begin{equation}
    \mathrm{p_t(t_d|t_d^{min},t_d^{max},d) \propto 
    \begin{cases}
    (t_d)^{-d} & \text{, for }  t_d^{min}<t_d<t_d^{max}, \\
    0 & \text{otherwise},
    \end{cases}}
\end{equation}
here, the delay time is defined as \( t_d = t_m - t_f \), where $t_m$ and $t_f$ correspond to the lookback times of the merger and the initial stellar formation, respectively. The merger rate of 1G BBHs at a given redshift $z$ can then be written as:
\begin{equation}
    \mathrm{R_{1G}(z) = R_0 \frac{\int_z^{\infty} p_t(t_d|t_d^{min},t_d^{max},d) R_{SFR}(z_f) \frac{dt}{dz_f} dz_f}{\int_0^{\infty} p_t(t_d|t_d^{min},t_d^{max},d) R_{SFR}(z_f) \frac{dt}{dz_f} dz_f}}.
\end{equation}

In this expression, the parameter $R_0$ refers to the local merger rate, which specifies the frequency of mergers at redshift $z = 0$. Based on the findings of \cite{LIGOScientific:2025pvj}, the estimated local BBH merger rate at $z = 0$ lies within the range $14$-$26 \, \mathrm{Gpc^{-3} \, yr^{-1}}$, quoted as the central 90\% credible interval. For the purpose of our analysis, we adopt a fiducial value of $R_0 = 20 \, \mathrm{Gpc^{-3} \, yr^{-1}}$ for the BBH population. The term $R_{\mathrm{SFR}}(z_f)$ denotes the star formation rate \cite{Madau:2014bja}, while $\frac{dt}{dz_f}$ corresponds to the Jacobian relating cosmic time and redshift.

\subsubsection{Hierarchical Mergers} 

Hierarchical mergers describe the process in which black holes produced in earlier coalescences subsequently merge with other black holes, giving rise to higher-generation remnants. In this work, we focus on second-generation (2G) black holes, formed directly through the merger of two first-generation (1G) black holes. Such 2G remnants can exceed the stellar-evolutionary mass cutoff, including the pair-instability gap. Unlike stellar-collapse black holes, they retain about $95\%$ of the total mass of their progenitors \cite{Pretorius:2005gq,Ossokine:2017dge}, and their spins typically cluster around $\chi \sim 0.7$ due to merger dynamics \cite{Scheel:2008rj,Campanelli:2006uy,Fishbach:2017dwv}.  

To model the 2G population, we assume a hierarchical formation channel where 2G mergers arise directly from 1G merger products. The corresponding merger rate is then obtained by convolving the 1G merger rate with a suitable delay-time distribution. This convolution introduces an additional timescale between successive mergers, which suppresses the overall 2G rate and shifts it toward lower redshifts. While the \texttt{BCO Phase Space} framework allows for more elaborate hierarchical scenarios, in this study we adopt a simplified delay-time prescription for clarity of presentation.

\begin{equation}
    \mathrm{R_{2G}(z) = R_0 \frac{\int_z^{\infty} p_t(t_d|t_d^{min},t_d^{max},d) R_{1G}(z_f) \frac{dt}{dz_f} dz_f}{\int_0^{\infty} p_t(t_d|t_d^{min},t_d^{max},d) R_{1G}(z_f) \frac{dt}{dz_f} dz_f}}.
\end{equation}

In general, delay times depend on properties of the host cluster, including its mass, radius, and escape velocity \cite{Stegmann:2022ruy,Chattopadhyay:2023pil,Antonini:2024het}. For the purposes of this work, we employ a simplified prescription that captures the essential features of dynamical mergers. This choice is motivated by the limited number of gravitational-wave detections available so far, which do not yet provide sufficient statistical power to constrain the full parameter space of cluster models. While detailed cluster simulations remain crucial for a comprehensive description, our simplified approach avoids introducing additional uncertainties that cannot currently be tested against observations.  

As a result, the estimated 2G merger rate naturally reflects the hierarchical character of black hole mergers, showing both suppression and a shift toward lower redshifts. Although the present model is sufficiently flexible to represent a variety of merger channels—including those influenced by cluster dynamics—the small number of high signal-to-noise ratio (SNR) detections means that more complex scenarios cannot yet be reliably constrained. Nevertheless, the \texttt{BCO Phase Space} framework can be straightforwardly extended to incorporate alternative merger prescriptions once richer observational datasets become available. For the 1G population we assume a characteristic delay time of $500 \, \mathrm{Myr}$, while 2G black holes are modeled with a longer delay of about $1 \, \mathrm{Gyr}$.

\subsubsection{Characterizing Mass and Spin of Astrophysical Origin Black Holes}
\label{sec:MassSpinABH}

The mass spectrum of astrophysical black holes is modeled as a mixture of two components: a Gaussian distribution and an exponential cutoff. This parametrization provides sufficient flexibility to describe a wide range of possible black hole mass distributions, and can be written as:

\begin{equation}
P(m) = 
    \begin{cases}
        \frac{1}{\sqrt{2\pi \sigma^2}} \exp\left(-\frac{(m - M_{\rm median})^2}{2 \sigma^2}\right), 
        & \qquad \text{if } m < M_{\rm median}, \\[10pt]
        \frac{1}{\sqrt{2\pi \sigma^2}} \exp\left(-\frac{(m - M_{\rm median})^2}{2 \sigma^2}\right) 
        \exp(-\alpha (m - M_{\rm median})), 
        &\qquad\text{if } m \geq M_{\rm median}.
    \end{cases}
    \label{eq:MassModel}
\end{equation}

\begin{figure}
\centering
\includegraphics[width=9cm, height=5.8cm]{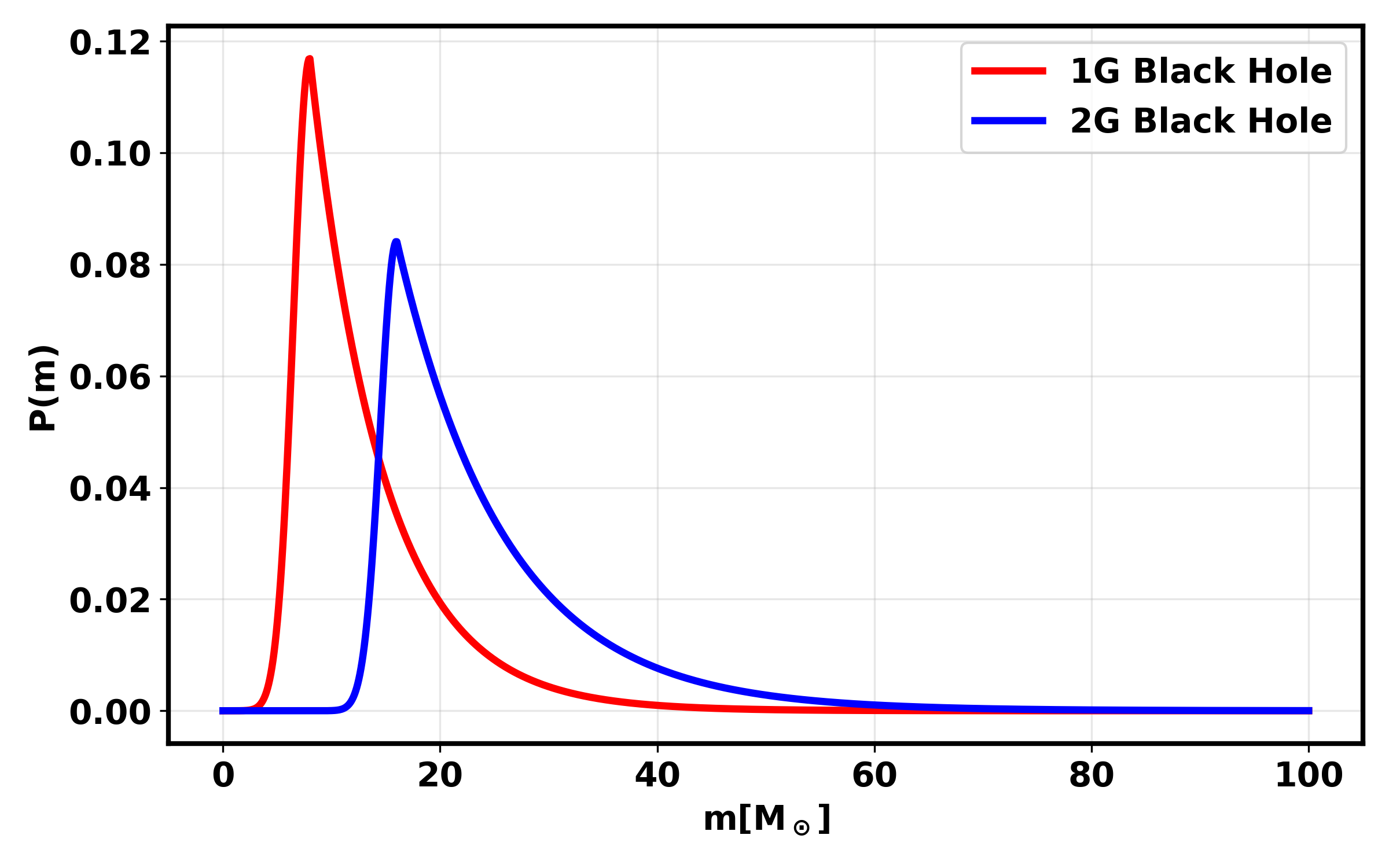}
\caption{The probability density functions for first-generation (1G) and second-generation (2G) black holes are illustrated. The 1G distribution (red curve) peaks at $8\,M_\odot$, whereas the 2G distribution (blue curve) shows a peak at $16\,M_\odot$.}
\label{fig:ABHMassModel}
\end{figure}

The combination of Gaussian and exponential components offers flexibility in modeling a wide range of astrophysical mass spectra, effectively capturing both the central peak and extended tails. The exponential cutoff in particular introduces the ability to reproduce asymmetric features, which are often present in observational data. The parameters $M_{\mathrm{median}}$, $\sigma$, and $\alpha$ provide intuitive control over the peak position, width, and decay rate, making it straightforward to align the model with measured distributions. Each of these parameters carries direct physical interpretation, which facilitates discussion of the astrophysical processes shaping the mass spectrum. Although this model is general enough to represent mass distributions expected from cluster-driven formation channels through appropriate parameter choices, the current catalog of GW detections remains too limited to place strong constraints on the full parameter space or on more complex alternatives.  

Figure~\ref{fig:ABHMassModel} shows illustrative probability density functions for two representative black hole populations: 1G and 2G remnants. The red curve corresponds to 1G black holes, peaking around $8\,M_\odot$, indicating that most of these sources are concentrated near this mass. The blue curve represents the 2G population, with a peak near $16\,M_\odot$, consistent with their systematically larger masses relative to 1G black holes.  

In the context of binary black hole systems, the dimensionless spin parameter $\chi$ is a key quantity for characterizing merger dynamics. For this study, we adopt a simplified description by modeling $\chi$ as a Gaussian-distributed variable,

\begin{equation}
    p(\chi) \propto \exp\left[-\frac{(\chi-\mu_\chi)^2}{2\sigma_\chi^2}\right],
\end{equation}

Here, $\mu_\chi$ and $\sigma_\chi$ represent the mean and standard deviation of the distribution, respectively. Although more elaborate spin–mass models have been suggested in the literature (e.g., incorporating effects of accretion history, mass-dependent growth, or hierarchical mergers), present GW observations do not yet place strong constraints on the detailed form of the spin distribution \cite{Stevenson:2022djs,Mandel:2018hfr,Fragione:2023kqv}. Spin measurements in particular remain far less precise than estimates of mass or redshift, making it challenging to discriminate between competing astrophysical models. For this reason, we adopt a Gaussian spin distribution as a simple yet flexible parametrization. This choice captures the main statistical variation in black hole spins while avoiding unnecessary complexity, with the understanding that the framework can be refined as future datasets provide tighter constraints.

\section{Reconstruction of Phase Space for Binary Black Holes}
\label{sec:PhaseSpaceGen}

To construct the phase space of astrophysical black holes, we divide the redshift interval $z \in [0,4]$ into 160 uniform bins, each with a width of $\Delta z = 0.025$. The total number of gravitational-wave events is then evaluated as

\begin{equation}
N_{\mathrm{GW}} = T_{\mathrm{obs}} \int_0^z \frac{dV_c}{dz}\frac{R(z)}{1+z}dz,
\label{eq:TotEvent}
\end{equation}

Here, $dV_c/dz$ represents the differential comoving volume element, $R(z)$ is the merger-rate density as a function of redshift, and $T_{\mathrm{obs}}$ denotes the observation time in years. The factor $(1+z)$ accounts for the cosmological time dilation of merger events.  

\begin{figure}
\centering
\includegraphics[width =10cm, height=8.5cm]{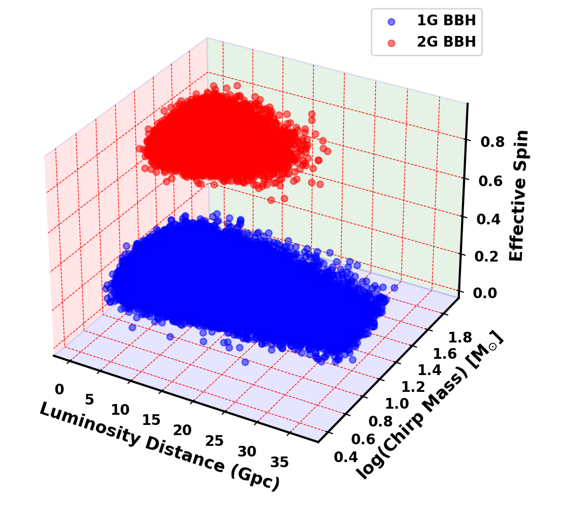}
\caption{Phase-space distribution of astrophysical black holes as a function of chirp mass, luminosity distance, and effective spin. Distinct regions are populated by first-generation black holes and by their hierarchical second-generation (2G) counterparts.}
\label{fig:PhaseSpace}
\end{figure}

For each formation channel, we simulate GW detections over a 32-month observational period. The black hole mass and spin values are drawn from their respective probability distributions using inverse-transform sampling based on the cumulative distribution function (CDF). From these realizations we compute the chirp mass and effective spin, and then map the population into phase space, as shown in Figure~\ref{fig:PhaseSpace}. The astrophysical black hole (ABH) mass spectrum is modeled using a Gaussian core with exponential tails (see Section~\ref{sec:MassSpinABH}). For the 1G population we adopt parameters $\mathrm{M_{median}} = 8\,M_\odot$, $\sigma = 1.5\,M_\odot$, and $\alpha = 0.15$, while for 2G remnants the chosen values are $\mathrm{M_{median}} = 16\,M_\odot$, $\sigma = 2.0\,M_\odot$, and $\alpha = 0.06$.

The spin distributions are also modeled using Gaussian profiles. For 1G black holes, the distribution is centered at $\chi = 0.2$, while 2G black holes are described by a Gaussian with mean $\chi = 0.7$ and standard deviation $\sigma = 0.1$. To account for astrophysical delays, we impose minimum merger times of $500\,\mathrm{Myr}$ for 1G binaries and $1\,\mathrm{Gyr}$ for 2G binaries. As shown in Figure~\ref{fig:PhaseSpace}, these distinct formation channels populate different regions of parameter space.

\begin{figure*}
\centering
\includegraphics[width=0.45\textwidth, height=5.5cm]{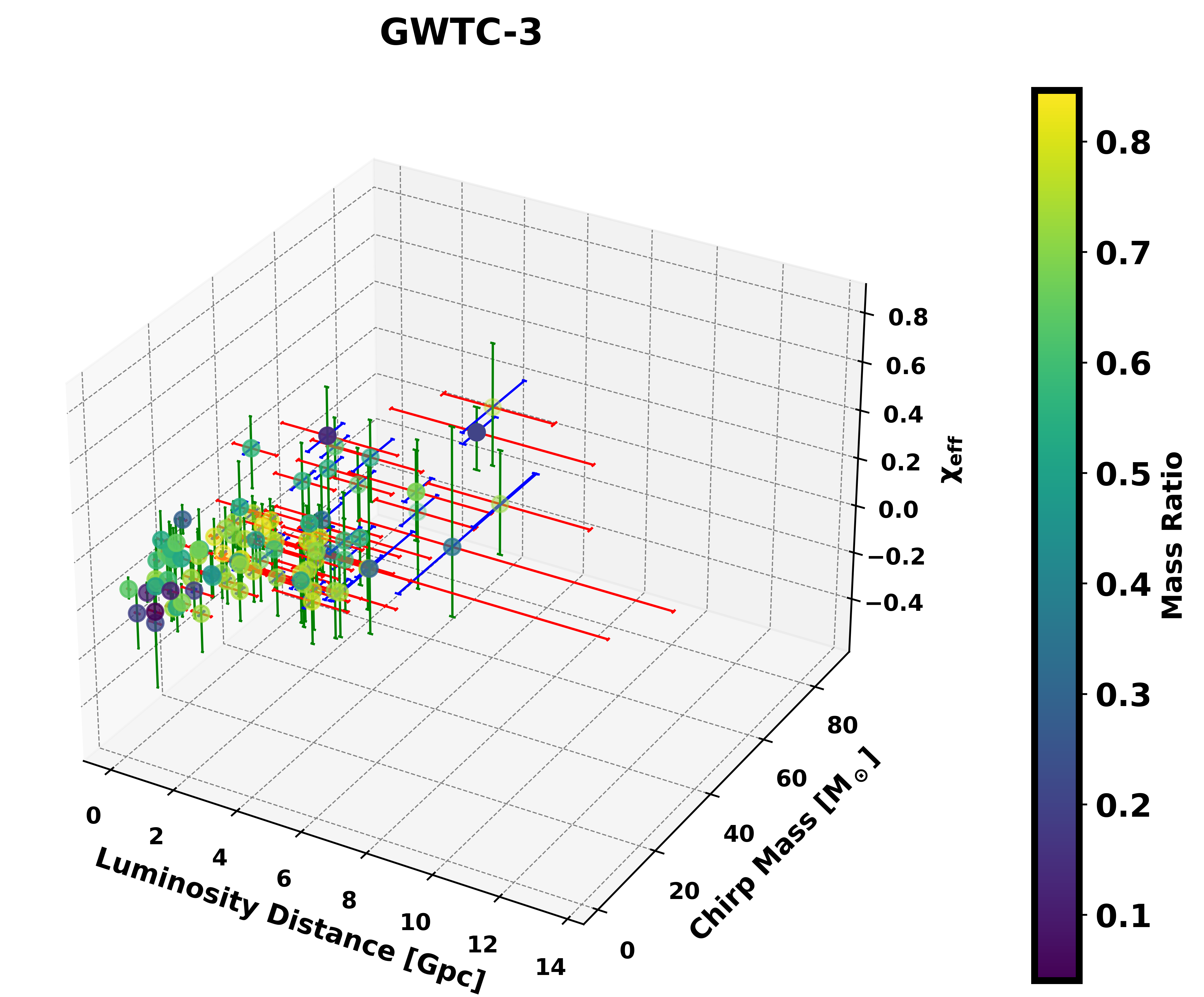}
\includegraphics[width=0.45\textwidth, height=5.5cm]{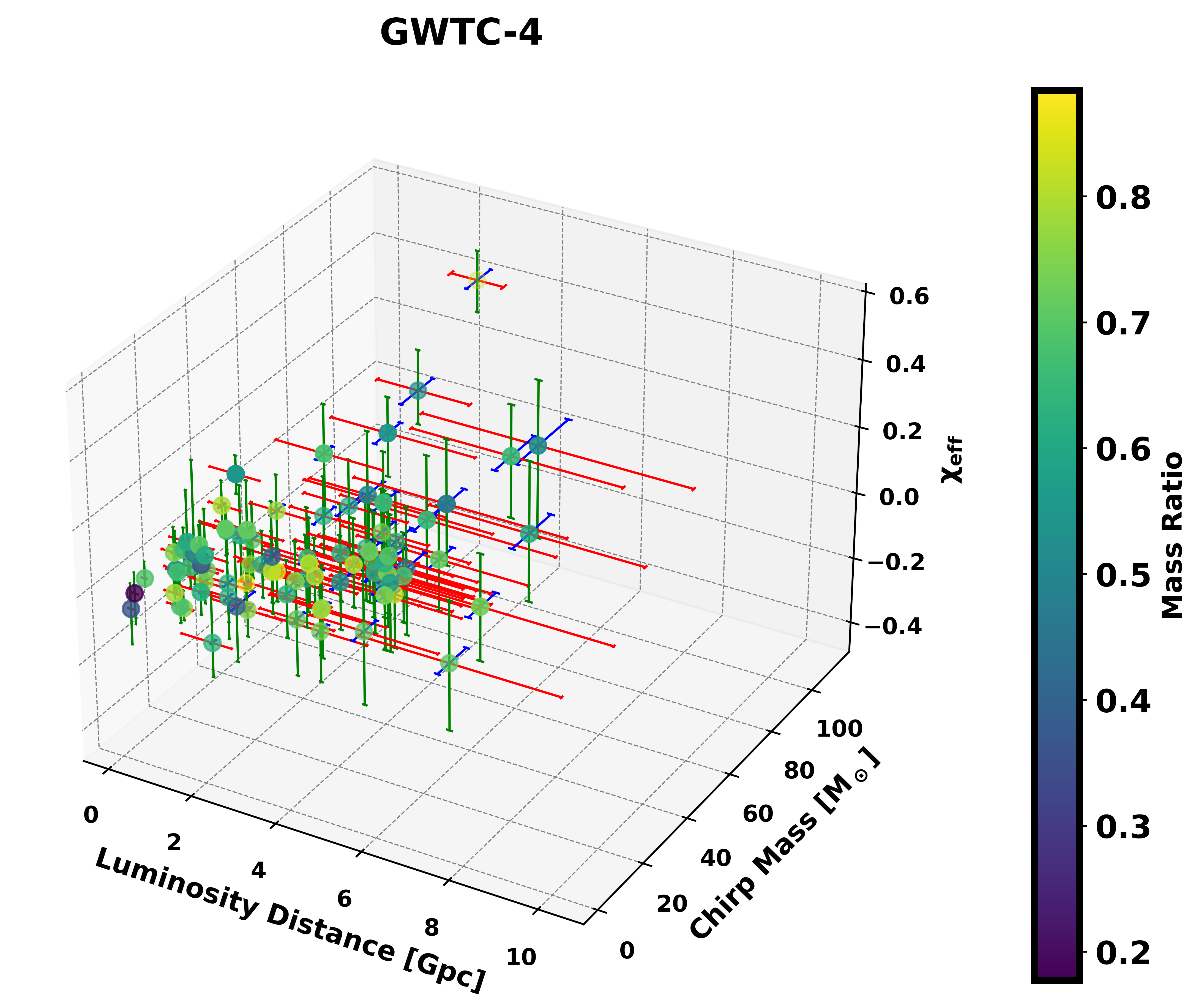}
\caption{The plot shows key source properties including Luminosity Distance (in Gpc), Chirp Mass (in solar masses), effective spin parameter ($\chi_{\mathrm{eff}}$), and Mass Ratio. The figure consists of two panels: the left panel shows events from the GWTC-3 catalog, while the right panel presents events from the GWTC-4 catalog. Data points represent median values of these parameters for detected binary mergers, with 1$\sigma$ error bars indicating the statistical uncertainties.}
\label{fig:EffParameterPlot}
\end{figure*}

\section{Identifying Formation Channels of Compact Objects from GWTC-4 catalog}
\label{sec:PhaseSpaceResults}

We investigate the possible formation channels of compact objects using the GW events reported by the LVK Collaboration, including all events up to GWTC-4, with the exception of GW170817.

To perform this study, we construct the phase space of GW events under two different cases. In the first case, we use the chirp mass ($M_c$), effective spin ($\chi_{eff}$), and luminosity distance ($D_L$) as the defining parameters, while in the second case we employ the component masses ($M_1$, $M_2$), individual spins ($\chi_1$, $\chi_2$), and luminosity distance. The observational data set are mapped into a three-dimensional parameter space for comparison with theoretical predictions corresponding to different formation scenarios. In Figure~\ref{fig:EffParameterPlot}, we show a three-dimensional phase-space plot of chirp mass, effective spin, and luminosity distance, including the $1\sigma$ uncertainties for all observed GW events. This framework allows us to systematically compare the observed distributions with the expectations from different astrophysical and primordial black hole formation channels.

\begin{figure*}
\centering
\includegraphics[width=0.48\textwidth, height=5.0cm]{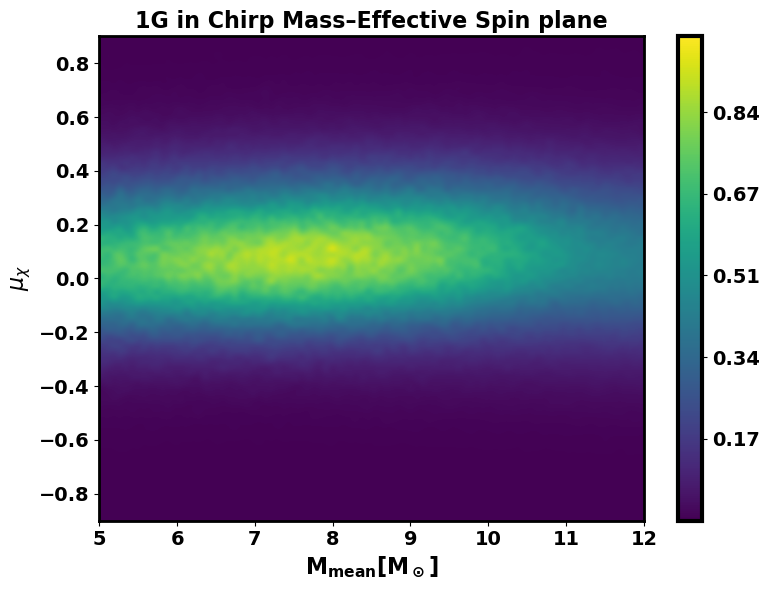}
\includegraphics[width=0.48\textwidth, height=5.0cm]{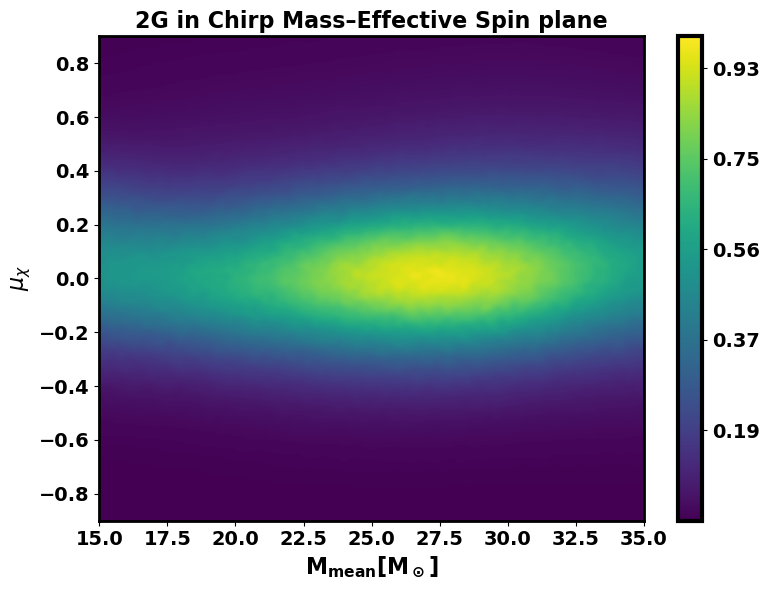}
\caption{Phase-space projections of binary black holes in terms of chirp mass \(M_c\), effective inspiral spin \(\chi_{\mathrm{eff}}\), and luminosity distance \(D_L\). The left panel corresponds to first-generation (1G) astrophysical black holes; the right panel corresponds to second-generation (2G) hierarchical mergers. The color bar in each panel indicates the probability weight associated with regions of the observable phase space after applying the detector selection function.}
\label{fig:EffPlaneProjection}
\end{figure*}

\begin{figure*}
\centering
\includegraphics[width=0.48\textwidth, height=5.0cm]{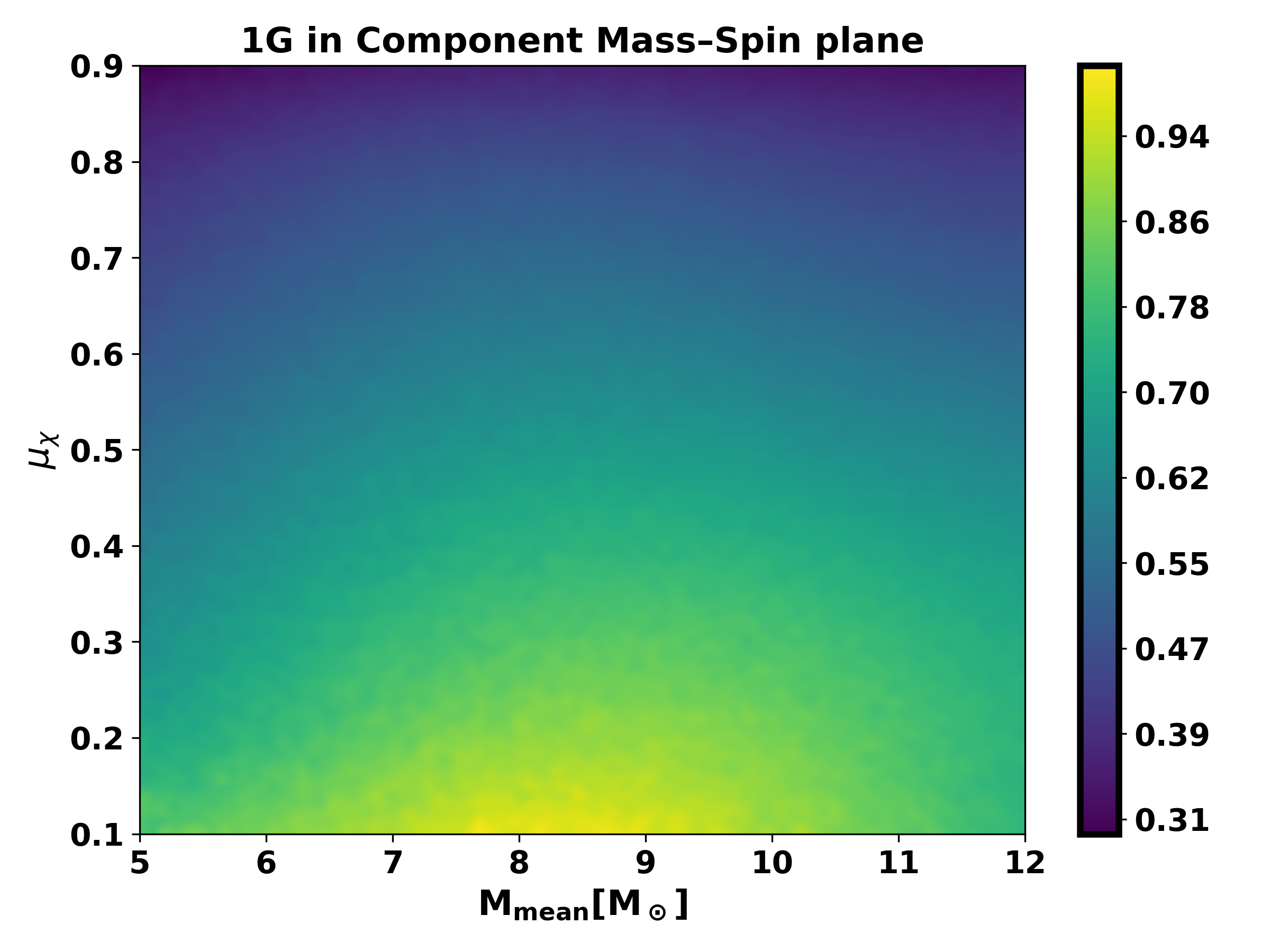}
\includegraphics[width=0.48\textwidth, height=5.0cm]{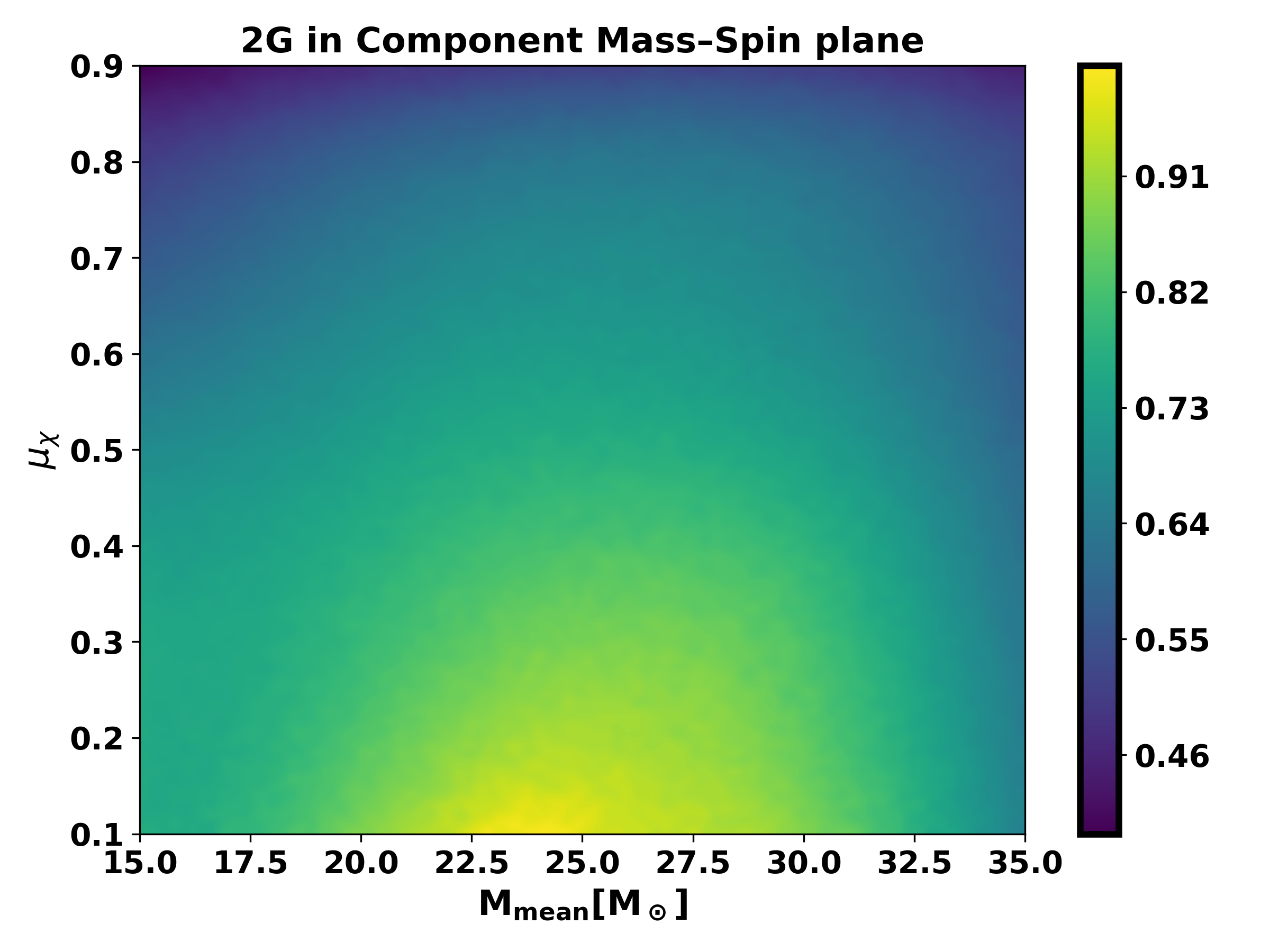}
\caption{Phase-space projections of binary black holes in terms of component masses \((M_1, M_2)\), individual spins \((\chi_1, \chi_2)\), and luminosity distance \(D_L\). The left panel corresponds to first-generation (1G) astrophysical black holes; the right panel corresponds to second-generation (2G) hierarchical mergers. The color bar in each panel indicates the probability weight associated with regions of the observable phase space after applying the detector selection function.}
\label{fig:MassSpinPlaneProjection}
\end{figure*}

We focus on two representative formation channels for compact objects: (i) first-generation (1G) astrophysical black holes formed from stellar collapse, and (ii) second-generation (2G) black holes produced through hierarchical mergers. While additional channels such as AGN-disk or dynamical cluster formation could also be incorporated, the limited number of high-confidence events in the LVK catalog motivates us to restrict our study to these two scenarios, which already capture the essential astrophysical uncertainties.

To explore the phase-space distributions associated with each channel, we vary only the parameters most directly linked to mass and spin. Specifically, we vary the mean of the Gaussian mass distribution and the mean of the Gaussian spin distribution. 

For 1G ABHs, the mass mean is allowed to vary between $5$ and $12\,M_{\odot}$, while for 2G ABHs it is varied between $15$ and $35\,M_{\odot}$. We note, however, that the underlying distribution is not truncated at these bounds; rather, the probability density for black hole mass extends to significantly higher values, as determined by the mass function model defined in Equation~\ref{eq:MassModel}. 

If we allowed the mass mean to go higher, for example up to $100,M_{\odot}$, our results would not be significantly affected because there are very few sources at such high masses, and their measurements are also much more uncertain. Consequently, these events contribute very little cumulative weight to the phase space and therefore do not materially influence our conclusions.

For both channels, the spin mean $\chi_{\mathrm{mean}}$ is varied across the full physical range, $0 \leq \chi_{\mathrm{mean}} \leq 1$. This parametrization provides a tractable yet flexible framework for capturing the dominant theoretical degeneracies, while retaining direct astrophysical interpretability in terms of the underlying black hole populations. These variations are designed to encompass uncertainties in dynamical channels where hierarchical mergers and mass segregation in dense clusters lead to broader mass distributions and more isotropic spin orientations \cite{Fragione:2023kqv,Rodriguez:2016,Zevin:2021,Mapelli:2020vfa,Antonini:2020xnd}.

This parameter space is employed as a prior in our likelihood analysis, enabling the generation of a diverse set of trajectories in the mass–spin–distance phase space. While the current GW catalogs contain only a few tens of high-SNR events, this framework is designed to remain flexible enough to capture richer astrophysical dependencies (e.g., metallicity or stellar evolution uncertainties) once larger event samples become available with future observing runs.

We project the theoretically generated trajectories for different formation channels onto the observed \texttt{BCO Phase Space}, which was constructed from GWTC-4 events except GW170817, as described above. This projection enables a direct comparison between model predictions and observational data within the common parameter space. 

To ensure that only detectable events are considered in our comparison, we apply a selection function that removes parameter combinations corresponding to sources that would fall below the LVK detection threshold. In this study, the selection function is computed using the exact matched-filter signal-to-noise ratio integrated over the detector noise power spectral density, which fully accounts for the frequency dependence of the waveform and detector sensitivity. Although detectability depends on several factors, it is primarily determined by the binary’s mass and luminosity distance, which therefore set the dominant shape of the selection function. Further details of the construction of the selection function are provided in Appendix ~\ref{sec:Framework}.

For the two formation channels \textbf{1G astrophysical black holes} and \textbf{2G hierarchical mergers} we generate a set of allowed trajectories in the observable phase space. These trajectories are constructed under the corresponding model assumptions and span the parameter grids described in the previous section. The observable phase space is defined in two complementary ways:  

\begin{itemize}
    \item \textbf{Case I (Chirp mass, effective spin, and distance):} using the chirp mass \((M_c)\), the effective inspiral spin parameter \((\chi_{\mathrm{eff}})\), and the luminosity distance \((D_L)\). This case exploits the fact that \(M_c\) and \(\chi_{\mathrm{eff}}\) are among the best-measured parameters in GW events, and provides a compact two-parameter description of the intrinsic binary properties along with distance. For the \textbf{1G BBH} channel, the probability distribution favors lower median chirp masses, around \(7\!-\!8\,M_\odot\), and low spins, with \(\chi_{\mathrm{eff}} \sim 0.1\), as shown in Figure~\ref{fig:EffPlaneProjection}. This trend arises because the majority of LVK events cluster at low masses with negligible effective spin. For the \textbf{2G BBH} channel, the results are less conclusive: regions of higher probability overlap significantly with those of the 1G BBH channel, leading to degeneracy. Nevertheless, when this overlap is disregarded, moderately larger chirp masses of about \(27\!-\!28\,M_\odot\) and effective spins of \(\chi_{\mathrm{eff}} \sim 0.1\) appear to be marginally favored.  

    \item \textbf{Case II (Component masses, individual spins, and distance):} using the component masses \((M_1, M_2)\), the individual spins \((\chi_1, \chi_2)\), and the luminosity distance \((D_L)\). Although individual spins are less well constrained, this formulation allows us to test the robustness of our classification framework when incorporating the full set of intrinsic binary parameters. When this case is considered, the qualitative trends remain broadly consistent with Case I, though additional features emerge due to the explicit treatment of poorly constrained spin parameters, as shown in Figure~\ref{fig:MassSpinPlaneProjection}. For 1G BBHs, the most probable region corresponds to masses of \(M \sim 8\,M_\odot\), with a spin peak around \(\chi \sim 0.1\), but with substantial contributions from higher spin values, reflecting the fact that spin measurements are largely unconstrained. For 2G BBHs, the most probable region corresponds to masses of \(M \sim 24\,M_\odot\), again with a spin peak near \(\chi \sim 0.1\) and similarly broad support at higher spins owing to weak spin constraints.  
\end{itemize}  

In both cases, the model trajectories are subjected to the detector selection function, which filters out sources below detectability thresholds. The resulting detectable distributions are then compared with the observed phase-space density of LVK events to quantify the degree of overlap for each formation channel. If we extend the priors to allow for higher masses, for example up to \(100\,M_\odot\), our results remain largely unaffected, since very few sources occupy this region of phase space and their parameter estimates are considerably more uncertain, contributing little statistical weight.

\begin{figure*}
\centering
\includegraphics[width=1.00\textwidth, height=20cm]{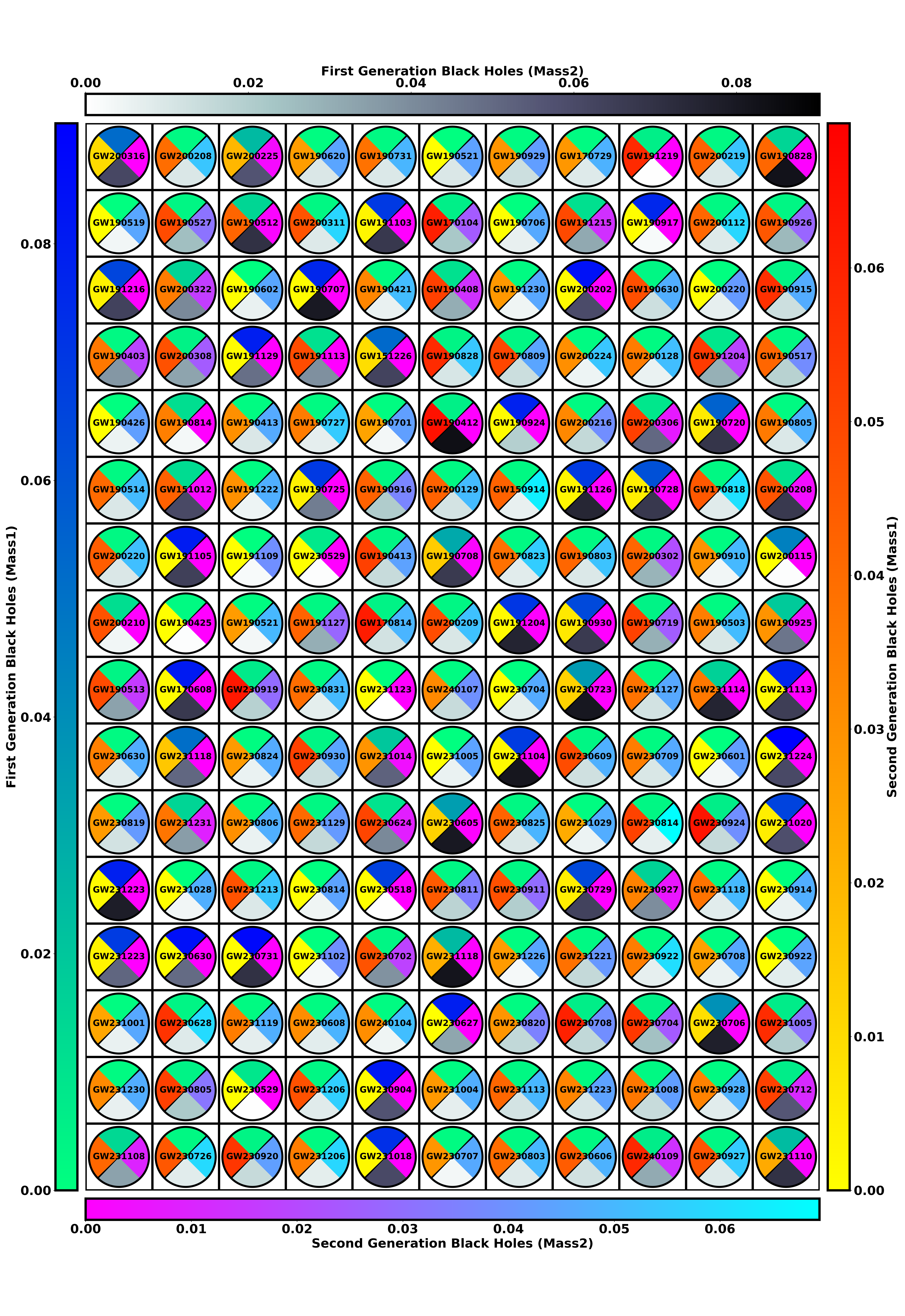}
\caption{This figure presents the formation channel probabilities for compact binary merger events from GWTC-4, excluding GW170817. For each event, the two pie-chart halves represent the two binary components: the upper half corresponds to the primary component, and the lower half to the secondary. Each pie wedge shows the relative probability that the component originates from either a first-generation (1G) or second-generation (2G) black hole formation channel, inferred from mass and spin measurements. These probabilities indicate the likelihood of each formation channel across the population of events, rather than probabilities normalized within individual events. Therefore, the wedge values should be interpreted as channel-specific relative likelihoods and are not expected to sum to unity for a single event.}
\label{fig:ProbabilityFormation}
\end{figure*}

Building on this framework, we can directly employ the phase-space overlap to classify individual GW events by their most likely formation channel \cite{Mould:2023ift,Kimball:2019mfs,Andrews:2020pjg}. Such classification offers critical insights into the astrophysical origins of compact binary coalescences. In this work, we employ a probabilistic framework that simultaneously incorporates the masses and spins of the compact objects, as these parameters together encode important information about their formation history. For each GW event, we calculate a probability weight quantifying the consistency of the source with the two formation channels: 1G black holes and 2G black holes. This weighting is derived by comparing the event’s posterior distributions-obtained via GW parameter estimation—with the theoretical mass-spin distributions predicted for each channel.

Formally, the weight assigned to a specific channel is defined by the joint mass-spin distribution as

\begin{equation}
w_{\text{channel}}^{(M,\chi)} = 
\frac{\int \int P_{\text{event}}(M, \chi)\, P_{\text{channel}}(M, \chi)\, dM\, d\chi}
{\int \int P_{\text{channel}}(M, \chi)\, dM\, d\chi},
\end{equation}

where \(P_{\text{event}}(M, \chi)\) represents the joint posterior probability density of the source’s mass and spin parameters inferred from GW data, and \(P_{\text{channel}}(M, \chi)\) is the normalized mass-spin distribution predicted by the astrophysical scenario.

These channel-specific distributions are constructed from marginalized projections illustrated in Figure~\ref{fig:MassSpinPlaneProjection}, which presents the mass and spin probability distributions for each formation channel. This visual summary, shown in Figure ~\ref{fig:ProbabilityFormation}, aids in interpreting the probability weights assigned to each event. Alternative classification strategies may utilize parameters such as the chirp mass and effective spin, as demonstrated in \cite{Afroz:2024fzp}.

In this study, redshift-dependent weights are not incorporated because the analyzed events are confined to a relatively narrow redshift range, where redshift alone provides limited discriminatory power. Looking forward, incorporating redshift priors with third-generation detectors is expected to significantly enhance channel discrimination and help resolve degeneracies between formation scenarios.

\begin{figure*}
\centering
\includegraphics[width=0.48\textwidth, height=5.0cm]{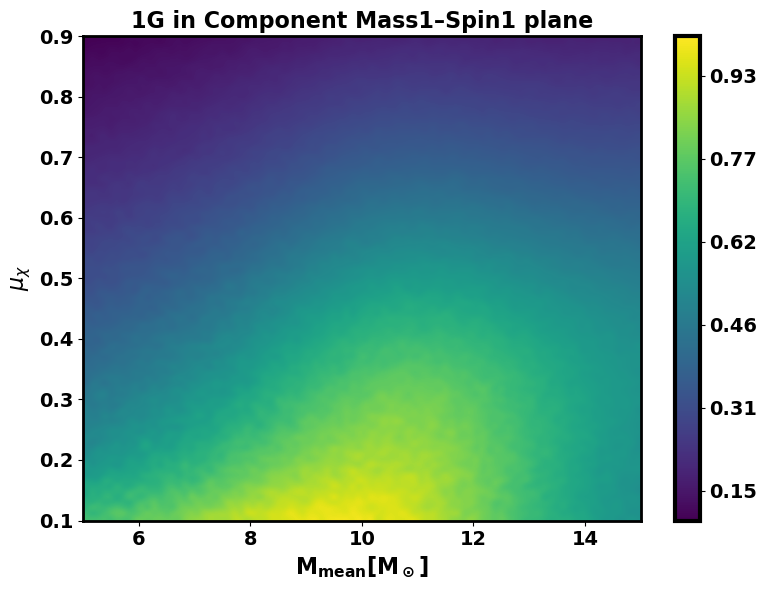}
\includegraphics[width=0.48\textwidth, height=5.0cm]{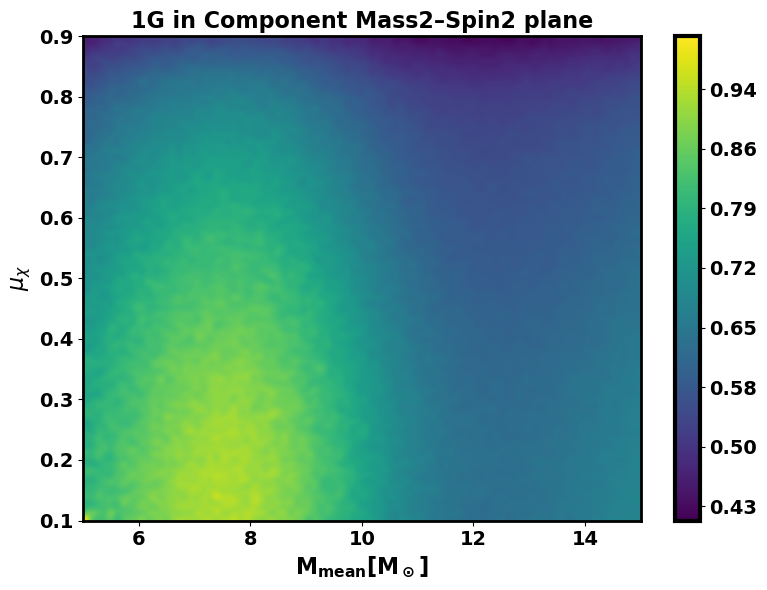}
\caption{Phase-space projections of binary black holes are shown separately for first-generation black holes in the primary component \((M_1, \chi_1, D_L)\) and the secondary component \((M_2, \chi_2, D_L)\). The left panel corresponds to the primary component, while the right panel corresponds to the secondary component. In each panel, the color bar indicates the probability weight associated with regions of the observable phase space. These joint distributions provide insights into the underlying physical processes, including clear signatures of the PISN mass gap, which imposes a fundamental boundary on black hole masses formed through stellar evolution.}
\label{fig:mass_spin_projections1G}
\end{figure*}

\begin{figure*}
\centering
\includegraphics[width=0.48\textwidth, height=5.0cm]{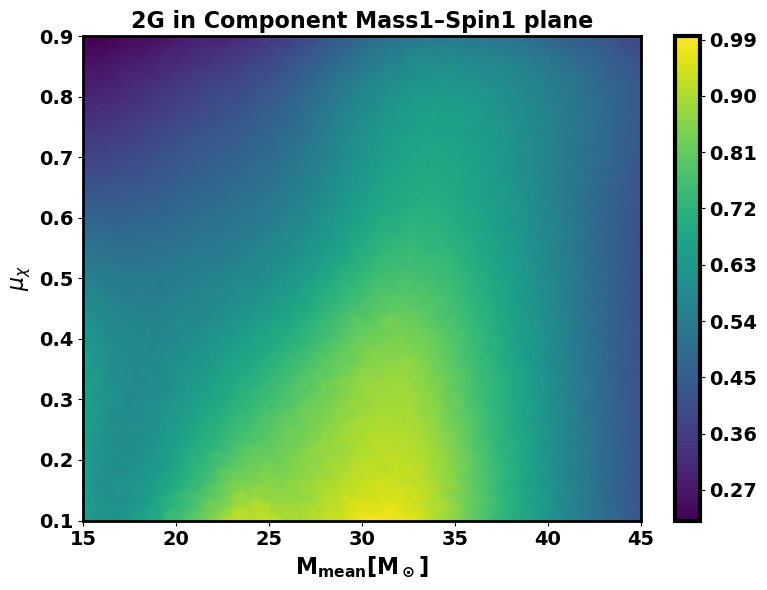}
\includegraphics[width=0.48\textwidth, height=5.0cm]{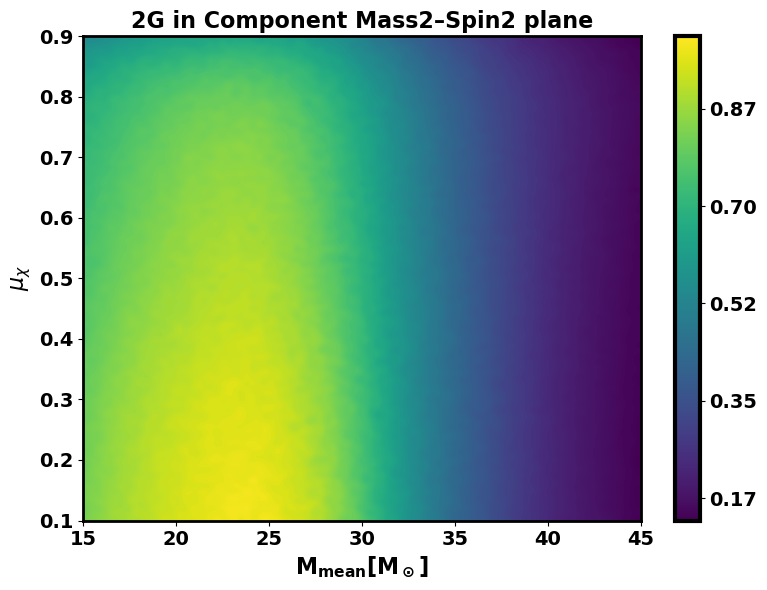}
\caption{Phase-space projections of binary black holes are shown separately for Second-generation black holes in the primary component \((M_1, \chi_1, D_L)\) and the secondary component \((M_2, \chi_2, D_L)\). The left panel corresponds to the primary component, while the right panel corresponds to the secondary component. In each panel, the color bar indicates the probability weight associated with regions of the observable phase space.}
\label{fig:mass_spin_projections2G}
\end{figure*}

\section{Phase-Space Evidence for the Pair-Instability Mass Gap}
\label{sec:PISNMassScale}
Building on the phase-space projections and event classification outlined above, we identify striking evidence for the presence of the PISN mass gap in the observed black hole mass distribution. The PISN gap represents a fundamental astrophysical boundary that shapes the distribution of binary black hole masses. The PISN phenomenon arises from the explosive disruption of massive stellar cores due to pair production instabilities, which theoretically prevents black hole formation within a specific mass range, typically spanning \(\sim 45 - 50\, M_{\odot}\) up to \(\sim 120\, M_{\odot}\) \cite{Woosley:2021xba,Marchant:2020haw}. This creates a characteristic “mass gap” in the black hole birth function that can be probed directly via GW observations \cite{Costa:2022aka, Fishbach:2017zga, Talbot:2018cva, Farmer:2020xne, Antonini:2025zzw, Li:2023yyt, Tong:2025wpz}.

To further elucidate this critical feature, we analyze the primary \((M_1, \chi_1)\) and secondary \((M_2, \chi_2)\) mass-spin parameter space projections for first-generation (1G) and second-generation (2G) black holes, as presented in Figures~\ref{fig:mass_spin_projections1G} and~\ref{fig:mass_spin_projections2G}.

Our study of the \(M_1-\chi_1\) and \(M_2-\chi_2\) distributions reveals a pronounced decline in the 1G population near the PISN cutoff, which we quantitatively identify as the mass where the cumulative probability density \(P(m)\) reaches 99.7\%. This corresponds to the 3$\sigma$ confidence level, a standard criterion in statistical inference. Adopting this threshold ensures that the definition of the cutoff is statistically rigorous, minimizes sensitivity to statistical noise or individual outliers in the high-mass tail, and allows for consistent comparison with other studies that employ similar significance levels.

\begin{figure*}
\centering
\includegraphics[width=0.48\textwidth, height=4.0cm]{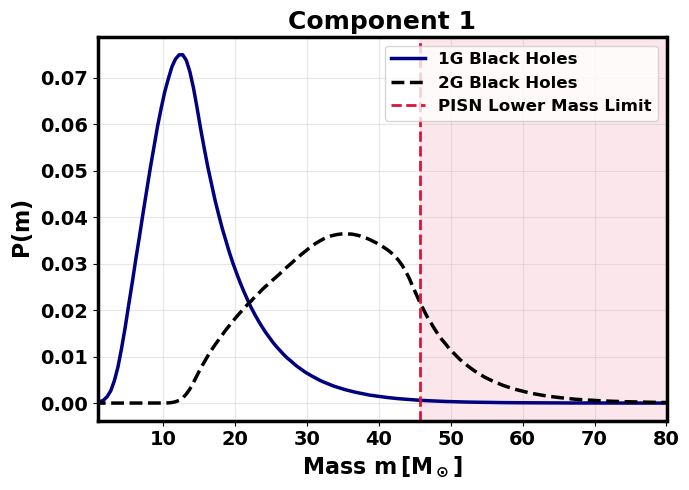}
\includegraphics[width=0.48\textwidth, height=4.0cm]{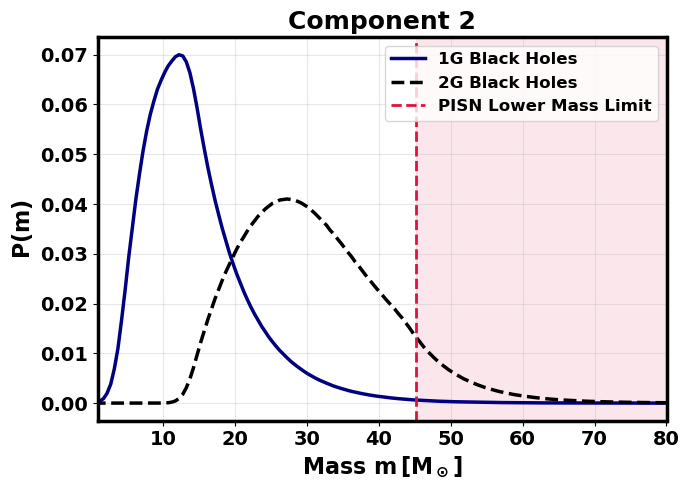}
\caption{Probability density functions of black hole masses reconstructed from weighted phase-space samples for first-generation (1G) and second-generation (2G) binaries. The left panel corresponds to component~1 \((M_1, \chi_1, D_L)\), and the right panel to component~2 \((M_2, \chi_2, D_L)\). The blue curves denote the mass distributions inferred from 1G binaries, while the black dashed curves represent the 2G cases from the Binary Compact Object (BCO) phase-space analysis. The vertical red dashed line marks the onset of the pair-instability supernova (PISN) gap at \( \sim 45\,M_\odot\), above which black hole formation from 1G progenitors is strongly suppressed. Although some support for black holes beyond this threshold appears in the 2G distribution, such cases remain significantly suppressed and typically originate from higher-mass, high-spin binaries. The shaded region highlights the PISN gap, reflecting the astrophysical boundary imprinted on the black hole mass spectrum.}
\label{fig:PISNMASS}
\end{figure*}

This cutoff occurs at approximately \(45.7\,M_{\odot}\) for the primary component and \(45.25\,M_{\odot}\) for the secondary (Figure~\ref{fig:PISNMASS}), reaffirming the physical limit on black hole mass set by isolated stellar evolution. Importantly, the spin distribution of 1G black holes tends to be narrowly clustered at low spin magnitudes, reflecting their formation through relatively undisturbed stellar collapse. By contrast, the 2G black hole population exhibits both an extended mass distribution beyond the PISN threshold and a broader range of spins, including significantly higher spin values arising from angular momentum acquired in hierarchical mergers. This combined characterization of mass and spin properties enables a robust segregation of 1G and 2G populations and underscores the astrophysical relevance of the PISN mass gap. These findings correspond closely to the schematic illustration shown in Figure~\ref{fig:Motivation}, where the sharp boundary in mass and spin phase space clearly separates the low-spin, lower-mass 1G black holes from the broader, higher-spin, and higher-mass 2G black hole population. This value of PISN lower mass gap is in agreement with the analysis from GWTC-3 and GWTC-4, which has shown a tentative evidence \cite{Karathanasis:2022rtr, Antonini:2025zzw, Tong:2025wpz}. 

To construct the underlying mass distribution from these phase-space weights, we generate large ensembles of samples drawn from mass and spin generative functions parameterized by key parameters such as the median mass of the distributions \(M_{\mathrm{median}}\), the standard deviation of the Gaussian component of the mass distribution \(\sigma\), the slope parameter of the mass distributions \(\alpha\), the mean of the spin distributions \(\mu_{\chi}\), and the variance of the spin distributions \(\sigma_{\chi}\), all indexed on the phase-space grid. Each sample is assigned a weight proportional to the phase-space probability density interpolated at the corresponding grid point. By aggregating these weighted samples across the entire parameter space, we build composite weighted histograms for the mass and spin distributions. These histograms are then normalized to yield probability density functions that inherently incorporate the full correlation and uncertainty structure encoded in the phase-space model. This sampling and weighting procedure provides an accurate means to translate complex model trajectories and phase-space weights into tangible mass distributions. In this study, we do not analyze the higher-mass cutoff of the PISN gap, since the number of observed high-mass events is very small and their measurements carry large uncertainties, so their cumulative contribution to the phase space is negligible.

This focused study highlights the utility of component-wise phase-space projections not only for classification purposes but also as a powerful tool to uncover and characterize critical astrophysical phenomena such as the PISN mass gap. It should also be noted that connecting the observed PISN gap to stellar evolution models involves additional astrophysical uncertainties, such as delay time distribution,  parent star’s metallicity and the dependence of stellar winds on the parent star metallicity \cite{Farmer:2019jed,Woosley:2021xba, Mukherjee:2021rtw}. As a result the inference of nuclear reaction rates directly from the current observation of the PISN is astrophysical model dependent and not robust.

\section{Conclusions}
\label{sec:conclusion}

In this study, we applied our binary compact object phase-space framework to the LVK catalog up to GWTC-4, with the aim of disentangling different compact-binary formation pathways. By analyzing the distribution of masses and spins across events, we find compelling evidence for the imprint of the PISN mass gap: the first-generation black hole population is truncated at $\sim 45.5 M_\odot$, while second-generation mergers populate the higher-mass regime beyond this boundary. This clear separation provides direct observational support for the role of the PISN limit in shaping the black hole mass spectrum. Importantly, the robustness of the PISN signature across parameterizations underscores the power of this approach to reveal astrophysical boundaries within GW data. An additional advantage of the phase-space formulation is its ability to incorporate the redshift dimension on the same footing as masses and spins. This allows us to probe potential redshift evolution of different formation channels, in contrast to approaches that rely solely on one-dimensional parameter distributions. In this way, the framework provides a more complete picture of how the relative contributions of different compact-binary populations may change across cosmic time. Furthermore it captures the overlap between formation channels, assigning probabilistic support across scenarios and highlighting current observational uncertainties.

Looking ahead, the method is naturally extendable to more detailed formation models and additional observables such as eccentricity, tidal effects, and electromagnetic counterparts. With the sensitivity of next-generation GW detectors, this framework will enable sharper distinctions between formation pathways and may uncover new classes of compact objects.

By demonstrating both the observational imprint of the PISN mass gap and the utility of phase-space analyses, our results provide a foundation for future studies of compact-binary evolution at the interface of stellar astrophysics and GW astronomy. As GW astronomy enters an era of increasingly rich datasets, our phase-space framework provides a powerful tool for unraveling the evolutionary histories of compact binaries. Beyond deepening our understanding of known astrophysical processes, it holds the potential to uncover new classes of compact objects, offering profound insights into stellar evolution and fundamental physics.

\section*{Acknowledgments}
The authors express their gratitude to Filippo Santoliquido for reviewing the manuscript and providing useful comments as a part of the LIGO publication policy. This work is part of the \texttt{⟨data|theory⟩ Universe-Lab}, supported by TIFR and the Department of Atomic Energy, Government of India. This research is supported by the Prime Minister Early Career Research Award, Anusandhan National Research Foundation, Government of India. The authors express gratitude to the system administrator of the computer cluster of \texttt{⟨data|theory⟩ Universe-Lab} and the TIFR computer center HPC facility for computing resources. Special thanks to the LIGO-Virgo-KAGRA Scientific Collaboration for providing noise curves. LIGO, funded by the U.S. National Science Foundation (NSF), and Virgo, supported by the French CNRS, Italian INFN, and Dutch Nikhef, along with contributions from Polish and Hungarian institutes. This collaborative effort is backed by the NSF’s LIGO Laboratory, a major facility fully funded by the National Science Foundation. The research leverages data and software from the Gravitational Wave Open Science Center, a service provided by LIGO Laboratory, the LIGO Scientific Collaboration, Virgo Collaboration, and KAGRA. Advanced LIGO's construction and operation receive support from STFC of the UK, Max-Planck Society (MPS), and the State of Niedersachsen/Germany, with additional backing from the Australian Research Council. Virgo, affiliated with the European Gravitational Observatory (EGO), secures funding through contributions from various European institutions. Meanwhile, KAGRA's construction and operation are funded by MEXT, JSPS, NRF, MSIT, AS, and MoST. This material is based upon work supported by NSF’s LIGO Laboratory which is a major facility fully funded by the National Science Foundation. We acknowledge the use of the following packages in this work: Numpy \cite{van2011numpy}, Scipy \cite{jones2001scipy}, Matplotlib \cite{hunter2007matplotlib}, and Astropy \cite{robitaille2013astropy}.

\section*{Data Availability}

The gravitational-wave catalogs employed in this study are publicly available on Zenodo as part of the LIGO–Virgo–KAGRA data releases: 
\href{https://zenodo.org/records/6513631}{GWTC-2.1}, \href{https://zenodo.org/records/5546663}{GWTC-3}, and \href{https://zenodo.org/records/16053484}{GWTC-4}.

\begin{appendices}

\section{Theoretical Framework of the Compact Object Phase Space Technique}
\label{sec:Framework}
The compact object phase space, first introduced in our earlier work \cite{Afroz:2024fzp, Afroz:2025efn}, provides a multidimensional representation that connects the observable properties of compact binaries inferred from GW measurements with theoretical predictions. The aim is to map observations onto model expectations in order to understand the physical processes driving the formation and evolution of compact objects. We define the phase space as an \(N\)-dimensional manifold, where each axis corresponds to a key observable. Distinct formation channels leave characteristic imprints in this space, allowing their systematic identification and comparison against data. Beyond validating existing models, the method also offers a model-agnostic avenue for discovering unexpected features, such as clustering, outliers, or unexplained regions that may signal novel sub-populations.

\subsection{Formulation in Terms of Effective Observables}

An alternative is to construct the phase space using combinations of parameters that are better constrained by GW data. The inspiral dynamics are dominantly governed by the \emph{chirp mass},
\begin{equation}
    \mathcal{M} = \frac{(m_1 m_2)^{3/5}}{(m_1+m_2)^{1/5}},
\end{equation}
which controls the leading-order frequency evolution of the inspiral waveform and is typically measured with percent-level precision even for moderate signal-to-noise ratio (SNR) events. Similarly, the spin contribution enters primarily through the \emph{effective inspiral spin},
\begin{equation}
    \chi_{\mathrm{eff}} = \frac{m_1 \chi_1 + m_2 \chi_2}{m_1+m_2},
\end{equation}
which determines whether the binary inspiral is accelerated (for aligned spins) or decelerated (for anti-aligned spins). This parameter is also relatively well constrained, although degeneracies with mass ratio and precessional effects remain. Together with the luminosity distance, these define a reduced phase space,
\begin{equation}
    \vec{X} = (\mathcal{M}, \chi_{\mathrm{eff}}, D_L).
\end{equation}

Each GW source again provides a posterior distribution,
\begin{equation}
    P_{\text{GW}}(\vec{X}) = P(\mathcal{M}, \chi_{\mathrm{eff}}, D_L),
\end{equation}
from which a normalized density can be constructed in the same manner as above. Because both \(\mathcal{M}\) and \(\chi_{\mathrm{eff}}\) are significantly better measured than their component counterparts, the resulting phase space representation has smaller statistical uncertainties and is therefore highly suitable for data-driven inference. The limitation of this approach is that some detailed astrophysical information, such as the mass ratio or spin misalignment angles, is effectively compressed into the effective parameters. This means that certain model-specific signatures may be partially washed out.

\subsection{Formulation in Terms of Component Quantities}

A natural starting point for constructing the phase space is to use the intrinsic component parameters of the binary, namely the mass \(M\), the dimensionless spin parameter \(\chi\), and the luminosity distance \(D_L\),
\begin{equation}
    \vec{X} = (M, \chi, D_L).
\end{equation}
These quantities are of direct astrophysical interest. The masses encode the outcome of stellar evolution and compact object formation, while the spin magnitudes and orientations carry information about processes such as tidal interactions, accretion, and possible dynamical capture in dense stellar environments. The luminosity distance connects the system to cosmology and to the distribution of sources across cosmic time.  

Each GW event provides a posterior probability distribution over this parameter space,
\begin{equation}
    P_{\text{GW}}(\vec{X}) = P(M, \chi, D_L),
\end{equation}
which encapsulates the uncertainties and correlations in the inference process. From this posterior we define a normalized density,
\begin{equation}
    Z(\vec{X}) = \frac{P_{\text{GW}}(\vec{X})}{\int P_{\text{GW}}(\vec{X}) \, d\vec{X}}, \qquad 
    \int Z(\vec{X}) \, d\vec{X} = 1,
\end{equation}
ensuring probabilistic consistency. For a population of detections, the combined density is given by the sum of the normalized contributions,
\begin{equation}
    Z_{\text{total}}(\vec{X}) = \sum_i Z_i(\vec{X}),
\end{equation}
which accumulates observational evidence across the entire catalog.

This component-based formulation has the important advantage that it maps directly to astrophysical models. Population synthesis simulations, for instance, often predict distributions of black hole masses and spin magnitudes, which can be compared straightforwardly to the observational phase space. However, current GW detectors measure these quantities with significant degeneracies, particularly between the total mass, mass ratio, and spin components. Consequently, while this representation is most faithful to the astrophysical variables of interest, it can be less robust for inference when uncertainties are large.

\subsection{Connecting to Theoretical Trajectories}

In both formulations, the cumulative observational distribution is compared to theoretical trajectories representing different formation channels. A model trajectory is expressed as
\begin{equation}
    T(\vec{X} \mid \{\lambda\}) = P_{\text{model}}(\vec{X} \mid \{\lambda\}),
\end{equation}
where \(\{\lambda\}\) are the model parameters. The likelihood that a given channel explains the data is quantified by the overlap,
\begin{equation}
    \mathrm{P_{\text{channel}}(\{\lambda\})} \propto \int Z_{\text{total}}(T(\vec{X} \mid \{\lambda\})) \, d\vec{X}.
\end{equation}
This integral measures the weighted consistency between theory and observation in the chosen parameterization.

Selection effects are incorporated through a detectability function,
\begin{equation}
    S(\vec{X}) = \Theta(\rho(\vec{X}) - \rho_{\text{th}}),
\end{equation}
with \(\rho(\vec{X})\) the signal-to-noise ratio (SNR) and \(\rho_{\text{th}}\) the detection threshold. The Heaviside function is defined as
\begin{equation}
    \Theta(x) =
    \begin{cases}
      1 & \text{if } x \geq 0, \\
      0 & \text{if } x < 0.
    \end{cases}
\end{equation}
This function ensures that only physically detectable regions of the phase space contribute to the likelihood. The effective likelihood is therefore
\begin{equation}
    \mathrm{P_{\text{channel}}(\{\lambda\})} \propto \int Z_{\text{total}}(T(S(\vec{X}) \mid \{\lambda\})) \, d\vec{X}.
\end{equation}

\subsection{Comparison of the Two Formulations}

The component-based representation \((M,\chi,D_L)\) has the virtue of being the most directly connected to astrophysical models and is particularly useful when the aim is to test population synthesis predictions or probe the detailed microphysics of compact binary formation. Its drawback lies in the relatively poor precision with which these quantities are currently measured, leading to broader posteriors and stronger degeneracies.  

The effective-parameter representation \((\mathcal{M}, \chi_{\mathrm{eff}}, D_L)\), by contrast, is more robust observationally. Because \(\mathcal{M}\) and \(\chi_{\mathrm{eff}}\) are among the best constrained GW observables, this phase space is less affected by noise and parameter correlations, making it powerful for model-agnostic statistical inference. Its main limitation is that certain astrophysical information is compressed, obscuring fine details such as spin tilts or asymmetric mass ratios.  

Taken together, the two approaches are complementary. The component-based formulation provides a natural bridge to astrophysical models, while the effective-parameter formulation maximizes the precision and robustness of GW inference. Employing both perspectives therefore yields the most comprehensive understanding of compact binary populations within the compact object phase space framework.

\end{appendices}

\bibliography{references}
\end{document}